\documentclass[
aps,%
12pt,%
final,%
notitlepage,%
oneside,%
onecolumn,%
nobibnotes,%
nofootinbib,% 
superscriptaddress,%
noshowpacs,%
centertags]%
{revtex4}

\usepackage{graphicx}
\begin{document}

\title{Heavy-flavour measurements in pp and Pb-Pb collisions with the 
       ALICE experiment at the CERN LHC}

\author{\firstname{Ralf}~\surname{Averbeck}}
\email{R.Averbeck@gsi.de}
\affiliation{%
ExtreMe Matter Institute EMMI and Research Division, 
GSI Helmholtzzentrum f\"ur Schwerionenforschung,
Darmstadt, Germany
}
\collaboration{for the ALICE Collaboration}
\noaffiliation

\begin{abstract}
The ALICE experiment at the CERN LHC has conducted first systematic studies
of heavy-flavour hadron production in pp collisions at $\sqrt{s} = 7$~TeV 
and in Pb-Pb collisions at $\sqrt{s_{\rm NN}} = 2.76$~TeV. 
In pp collisions the differential production cross sections of D mesons 
at mid-rapidity, as well as the cross sections for electrons and muons from 
semileptonic heavy-flavour hadron decays at mid- and forward-rapidity, 
respectively, have been measured. These data provide a crucial testing ground 
for perturbative QCD calculations in the new LHC energy regime.
In Pb-Pb collisions, the nuclear modification factor $R_{\rm AA}(p_{\rm t})$ has 
been measured for D mesons and for leptons from heavy-flavour decays, indicating
energy loss of heavy quarks in the partonic medium produced in Pb-Pb collisions
at the LHC. The strong interaction of charm quarks with this medium might also
generate a non-zero elliptic flow of D mesons as first studies of the
azimuthal anisotropy of D$^0$-meson production suggest.
\end{abstract}

\maketitle

\section{Introduction}
Particles carrying heavy flavour, {\it i.e.} charm or beauty quarks, are among 
the most interesting probes in contemporary particle and nuclear physics. Due
to the large quark mass, heavy-flavour production proceeds dominantly through 
hard partonic scattering processes in the earliest stage of hadronic collisions.
As such the measurement of heavy-flavour production in pp collisions at the
LHC provides a crucial testing ground for perturbative Quantum Chromodynamics
(pQCD), the theory of strong interactions, in a new high energy domain.

Furthermore, these measurements serve as a baseline for heavy-flavour studies
in Pb-Pb collisions at the LHC, where the heavy quarks propagate through and
interact with the hot and dense medium produced in the nuclear collisions.
The investigation of medium modifications of heavy-flavour observables will 
shed light on the properties of the medium and the nature of the parton-medium 
interactions. In particular, in-medium energy loss of heavy quarks is 
sensitive to the energy density and will reflect itself in the suppression
of heavy-flavour hadrons at high transverse momentum, \ensuremath{p_{\rm t}}. 
Such a suppression can be quantified using the nuclear modification factor $
R_{\rm AA}(p_{\rm t})$, which is the ratio of the \ensuremath{p_{\rm t}} 
distribution measured in nucleus-nucleus (in this case Pb-Pb) collisions at a 
given centrality and the corresponding distribution in pp collisions scaled by 
the average number of binary nucleon-nucleon collisions, 
$\langle N_{coll} \rangle$, in the selected centrality class of the 
nucleus-nucleus collisions:

\begin{equation}
R_{\rm AA}(p_{\rm t}) = \frac{1}{\langle N_{coll} \rangle}
                     \frac{dN_{\rm AA}/dp_{\rm t}}{dN_{pp}/dp_{\rm t}}
                   = \frac{1}{\langle T_{\rm AA} \rangle} 
                     \frac{dN_{\rm AA}/dp_{\rm t}}{d\sigma_{pp}/dp_{\rm t}}
\end{equation}

In this expression, the average nuclear overlap integral 
$\langle T_{\rm AA} \rangle$, which is obtained from a Glauber model calculation
of the collision geometry, provides a translation between cross sections as 
measured in pp collisions and particle multiplicities as measured in 
nucleus-nucleus collisions.

Energy loss of partons in a dense medium will lead to $R_{\rm AA}(p_{\rm t}) < 1$ 
for hadrons originating from the fragmentation of those partons. If energy loss 
proceeds mainly via induced gluon radiation in a coloured medium 
$R_{\rm AA}(p_{\rm t})$ should show a distinctive pattern. Gluons should lose more 
energy radiatively than quarks given the lower colour charge of quarks. 
For quarks, in addition, a mass hierarchy is expected. 
The 'dead-cone effect' should prohibit induced 
gluon radiation of a quark in a colour charged medium in a forward cone which 
increases in size with increasing quark mass~\cite{dead_cone}. Consequently, 
the nuclear modification factor of light hadrons, which predominantly originate
from hard scattered gluons or light quarks, should be larger than that of 
D mesons originating from charm quarks, which in turn should be larger than 
that of B mesons from beauty fragmentation: 
$R_{\rm AA}^\pi < R_{\rm AA}^D < R_{\rm AA}^B$. 
This hierarchy expected for radiative energy loss was not observed for 
electrons from heavy-flavour decays at RHIC~\cite{hfe_rhic1,hfe_rhic2},
where, up to now, electrons from charm and beauty decays could statistically be 
separated from each other in pp collisions only~\cite{b_phenix,b_star}. 
The direct measurement of $R_{\rm AA}^D$ in comparison with the nuclear 
modification factor of other probes is of prime importance to shed light 
on the mechanism of partonic energy loss.

Further insight can be gained from the measurement of the azimuthal anisotropy
of the charm hadron \ensuremath{p_{\rm t}} spectra with respect to the 
orientation of the reaction plane. 
In particular, the second Fourier coefficient, $v_2$, of this azimuthal 
distribution is expected to be sensitive to the degree of thermalisation of
charm quarks within the partonic medium. A non-zero $v_2$ would certainly be
indicative of a strong interaction of charm quarks with the medium.

The ALICE experiment was optimised for heavy-flavour measurements in
various decay channels (see Section 2). Measurements in pp collisions 
at $\sqrt{s} = 7$~TeV are presented in Section 3 and compared with pQCD 
calculations, which are in good agreement with the data. This justifies to
use pQCD to scale the pp data from $\sqrt{s} = 7$~TeV to 2.76~TeV in order
to obtain a reference for Pb-Pb collisions at that energy per nucleon-nucleon 
pair as described in Section 4. Results from heavy-flavour observables in 
Pb-Pb collisions at $\sqrt{s_{\rm NN}} = 2.76$~TeV are summarised in Section 5.

\section{Heavy-flavour measurements with ALICE}
The tracking and particle identification detectors of the ALICE experiment,
which is described in detail elsewhere~\cite{alice}, 
allow for excellent measurements of heavy-flavour production through the full 
reconstruction of hadronic D-meson decays at central rapidity as well as via 
semileptonic decays of charm and beauty both at central and forward rapidities. 

The tracking system consists of the silicon Inner Tracking System 
(ITS)~\cite{alice,its} and the Time Projection Chamber (TPC)~\cite{alice,tpc} 
located in a solenoidal magnetic field of 0.5~T. In the pseudorapidity range 
$|\eta| < 0.9$ tracks are reconstructed with a momentum resolution better than 
4\% for $p_{\rm t} < 20$~GeV/$c$. The distance of closest approach of tracks to 
the interaction vertex is measured with a resolution better than $75 \mu$m for 
$p_{\rm t} > 1$~GeV/$c$ in the plane transverse to the beam 
direction~\cite{d_in_pp}.

Charged hadrons are identified via their specific energy loss d$E$/d$x$ in the
TPC and a time-of-flight (TOF) measurement in the ALICE TOF detector. The
same techniques allow for a clean electron identification in the range
$p_{\rm t} < 6$~GeV/$c$. At higher \ensuremath{p_{\rm t}}, the Transition 
Radiation Detector (TRD) and, alternatively, the Electromagnetic Calorimeter 
(EMCal) are necessary to identify electrons. 
Muons are identified and their momenta are measured
in the forward muon spectrometer which covers the pseudorapidity range
$-4 < \eta < -2.5$.

The measurement of D mesons at mid-rapidity ($|y| < 0.5$) is based on the
selection of displaced-vertex topologies~\cite{d_in_pp}. Tracks are selected 
which originate from a secondary vertex consistent with a large decay length 
as expected from the D-meson lifetimes. In addition, a good alignment of the 
reconstructed D-meson momentum with the flight line from the collision vertex 
to the secondary vertex is required. Since all D-meson decay channels studied 
(the most important are ${\rm D}^0 \rightarrow {\rm K}^- \pi^+$, 
${\rm D}^+ \rightarrow {\rm K}^- \pi^+ \pi^+$, and
${\rm D}^{*+} \rightarrow {\rm D}^0 \pi^+$) involve a kaon as decay product
the identification of the charged kaon in the TPC and TOF detectors helps to
reduce the background. In an invariant mass analysis the raw D-meson yields 
are determined which, then, are corrected for geometrical acceptance and for
reconstruction and particle identification efficiency based on a detailed
simulation of the apparatus. Feed down from B-meson decays is estimated
from a fixed-order next-to-leading-log (FONLL) pQCD 
calculation~\cite{fonll} and it is subtracted. 

Electrons from heavy-flavour decays are measured at mid-rapidity ($|y| < 0.8$)
in a two step procedure. First, the inclusive electron spectrum is determined. 
Electron candidate tracks are identified with the TPC, TOF, and TRD detectors. 
The small remaining hadron background is estimated via fits of the TPC 
d$E$/d$x$ distribution in momentum slices and it is subtracted from the 
electron candidate spectra. After corrections for geometrical acceptance and 
tracking and particle identification efficiencies derived from simulations the 
inclusive electron spectrum is obtained. From this inclusive spectrum a 
cocktail of electrons from sources other than heavy-flavour decays is 
subtracted. The most important components of this background cocktail are 
Dalitz decays of the $\pi^0$ and $\eta$ meson and photon conversions in the 
beam pipe and in the first pixel layer of the ITS. In addition, decays of light
vector mesons ($\rho, \omega, \phi$), heavy quarkonia ($J/\psi, \Upsilon$), and
direct radiation from hard scattering processes are included. The cocktail input
is based on the $\pi^0$ and $\eta$ spectra measured by ALICE~\cite{pieta_alice},
the known material budget of the ALICE apparatus~\cite{koch_alice} 
(conversions), \ensuremath{m_{\rm t}}-scaling (light vector mesons), heavy 
quarkonia measurements from ALICE~\cite{jpsi_alice} and 
CMS~\cite{jpsi_cms,upsilon_cms}, and next-to-leading-order pQCD 
calculations~\cite{vogelsang1,vogelsang2} (direct radiation).

In a similar approach muons from heavy-flavour decays are measured at forward
rapidity~\cite{muon_pp}. From the inclusive muon spectrum three main background
sources are subtracted: muons from the decay of primary light hadrons, muons 
from the decay of hadrons produced in secondary interactions in the front 
absorber, and hadrons that punch through the front absorber. Punch through 
hadrons can be rejected requiring that muon candidate tracks match a track in 
the muon trigger system. The decay muon background dominates in the 
low-\ensuremath{p_{\rm t}} region and it can be subtracted using information 
from detailed simulations.

\begin{figure}
\includegraphics[width=0.3\textwidth]{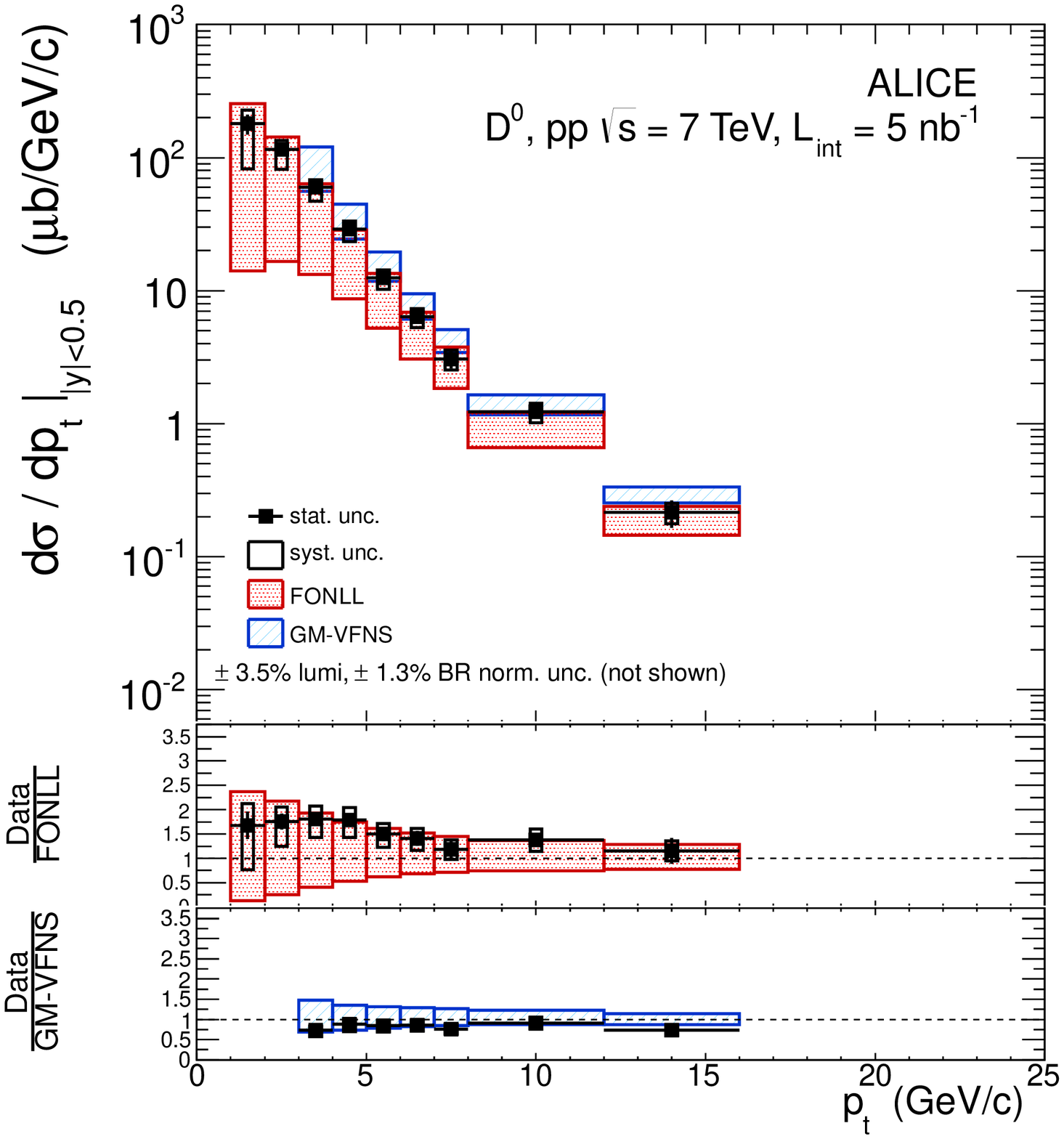}
\includegraphics[width=0.3\linewidth]{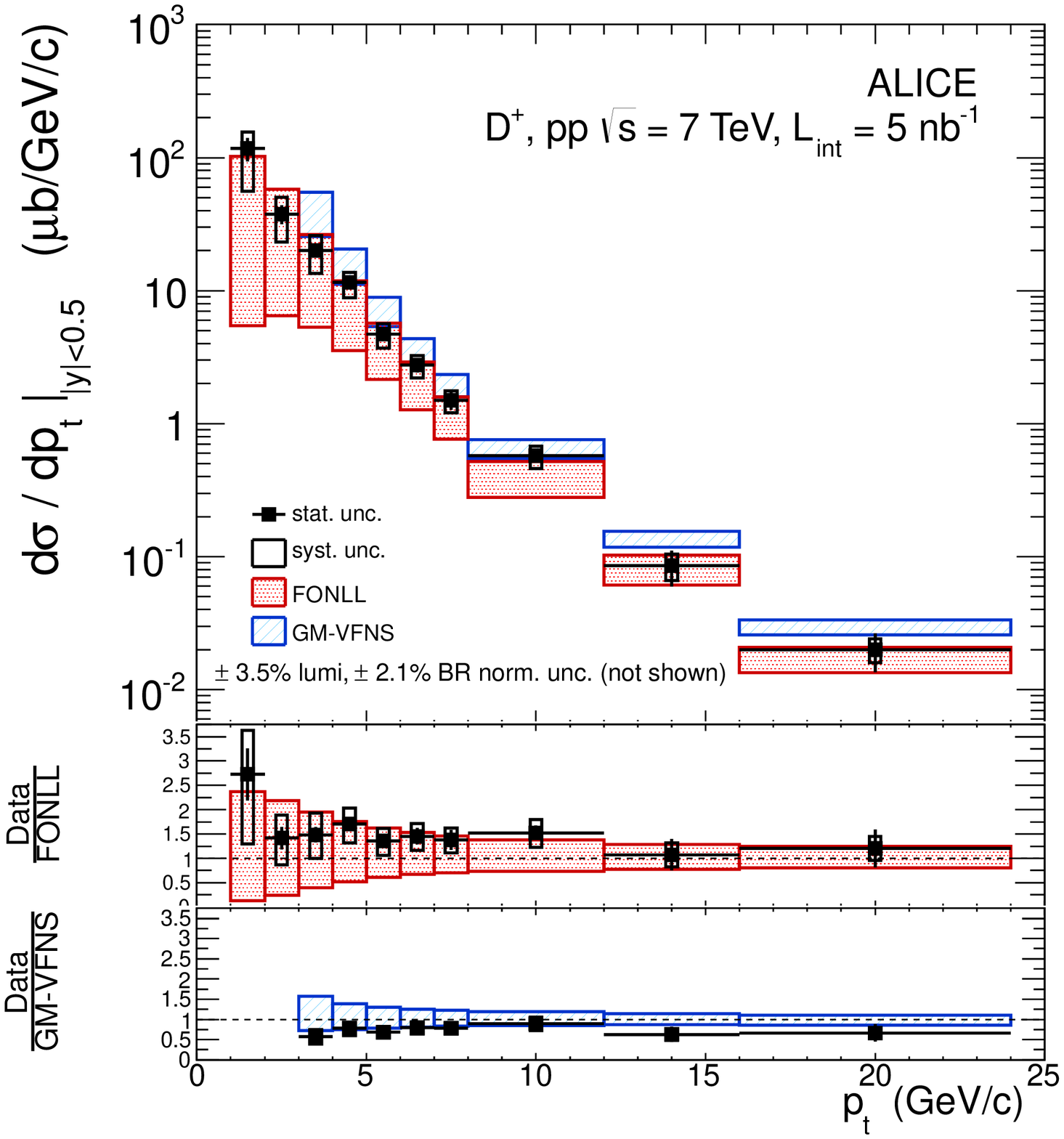}
\includegraphics[width=0.3\linewidth]{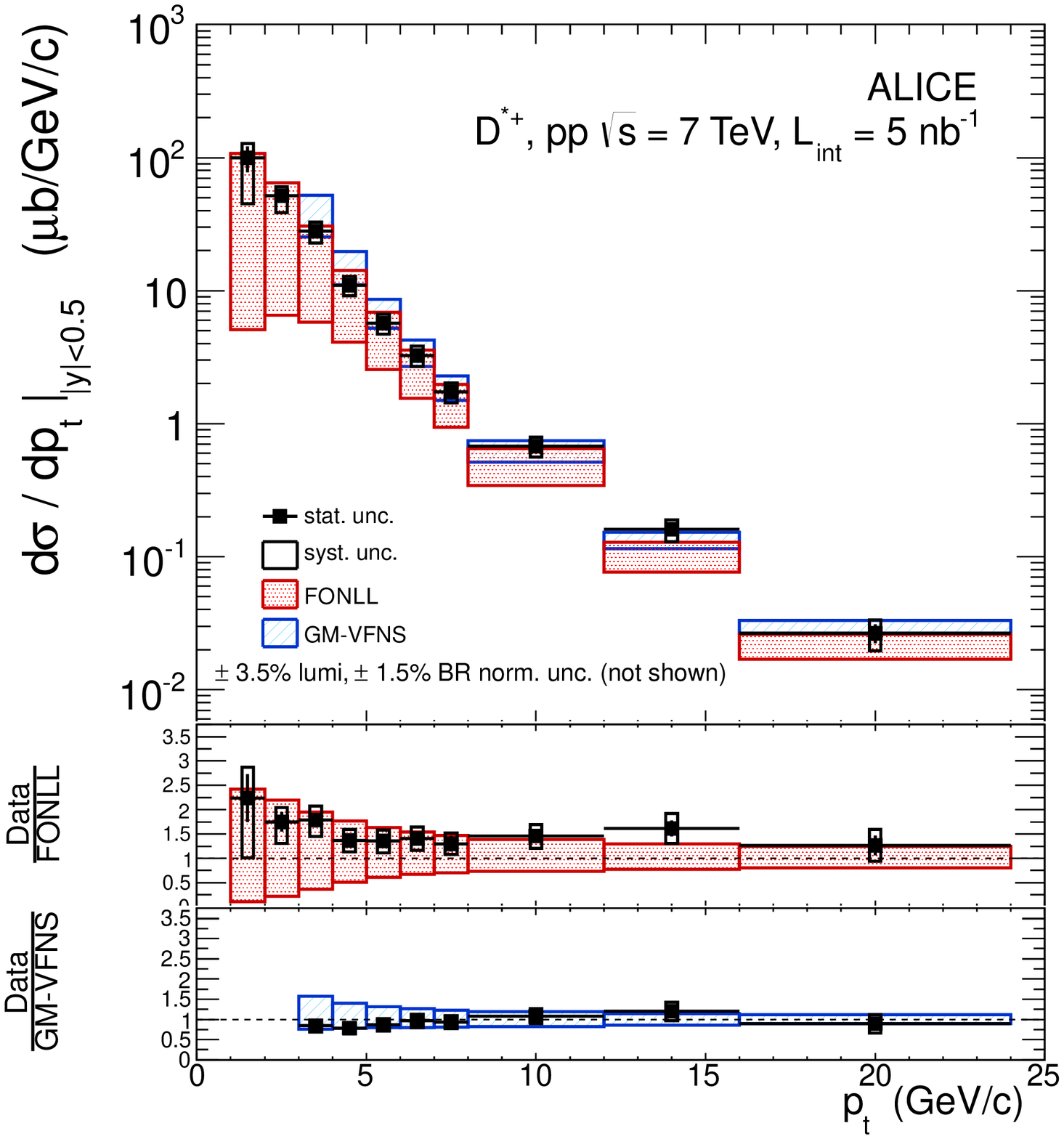}
\caption{${\rm D}^0$, ${\rm D}^+$, and ${\rm D}^{*+}$ 
         \ensuremath{p_{\rm t}}-differential production cross sections in pp 
         collisions at $\sqrt{s} = 7$~TeV~\cite{d_in_pp} in comparison with 
         pQCD calculations~\cite{fonll,gmvfns}.}
\label{fig:d_in_pp}
\end{figure}

The results presented here were obtained from data recorded in pp and Pb-Pb
collisions using minimum bias trigger selections. The minimum bias trigger
in pp collisions required the presence of signals in either of two 
scintillator hodoscopes (VZERO detectors), located in the forward and backward 
regions of the experiment, or in the silicon pixel detector (SPD) of the ITS.
In Pb-Pb collisions signals in both VZERO detectors and the SPD were required.
For the results presented here, 100-300 million (depending on the analysis) 
pp collisions and 17 million Pb-Pb collisions were analysed. For pp 
collisions production cross sections were normalised relative to the minimum
bias trigger cross section which was determined using a van der Meer 
scan~\cite{vdM}.
Yields in Pb-Pb collisions were measured in classes of collision centrality
which were defined based on a Glauber-model analysis of the sum of the
amplitudes measured in the VZERO detectors~\cite{centrality}.

\section{Heavy flavour in pp collisions at \mbox{$\sqrt{s} = 7$~TeV}}

The ${\rm D}^0$, ${\rm D}^+$, and ${\rm D}^{*+}$ 
\ensuremath{p_{\rm t}}-differential production cross sections measured at 
mid-rapidity ($|y| < 0.5$)~\cite{d_in_pp} are shown 
in Fig.~\ref{fig:d_in_pp}. Theoretical calculations based on 
next-to-leading-order pQCD calculations (FONLL~\cite{fonll} and 
GM-VFNS~\cite{gmvfns}) are in agreement with the data within substantial 
experimental and theoretical systematic uncertainties. 

\begin{figure}
\includegraphics[width=0.45\textwidth]{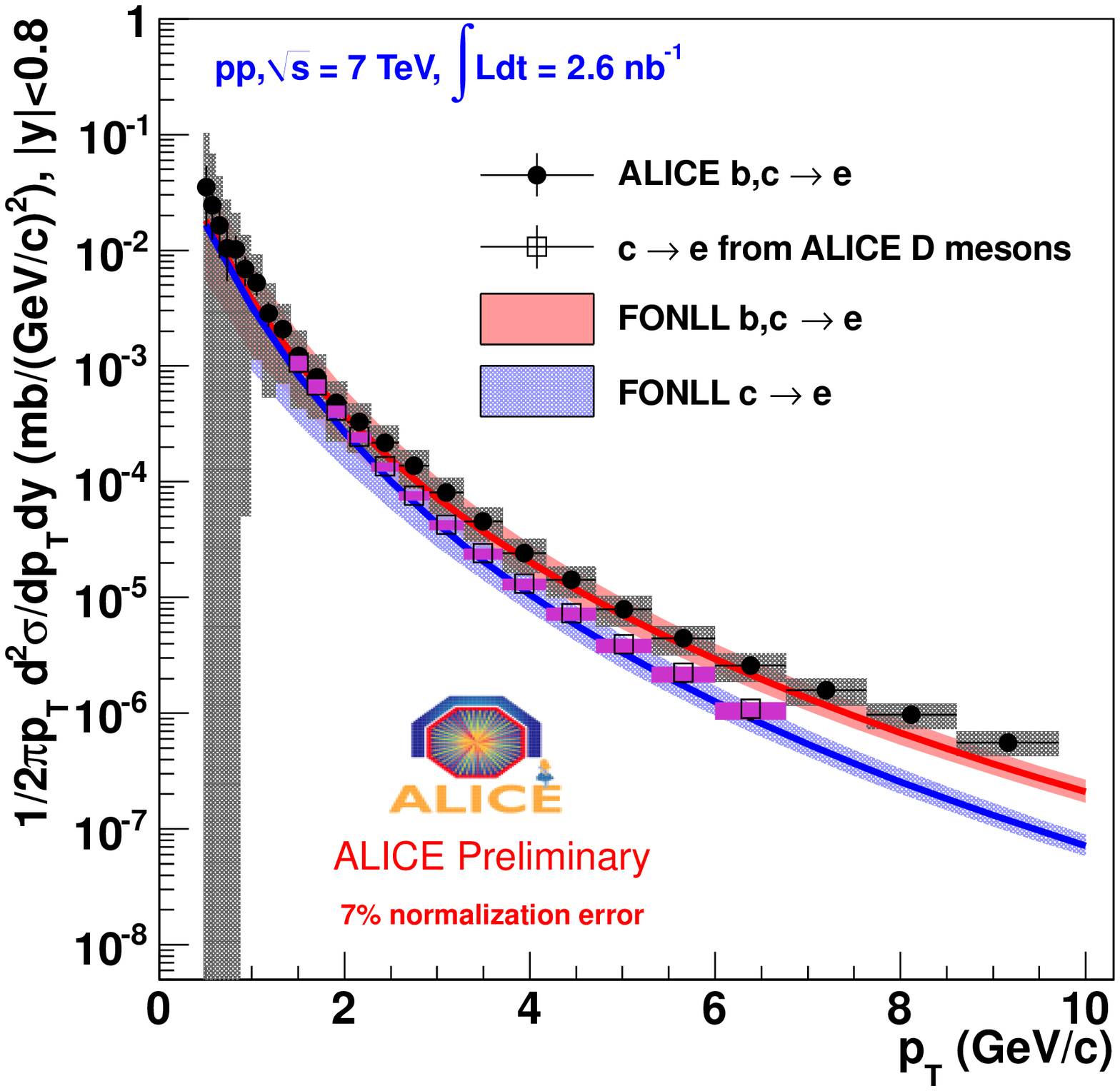}
\includegraphics[width=0.45\linewidth]{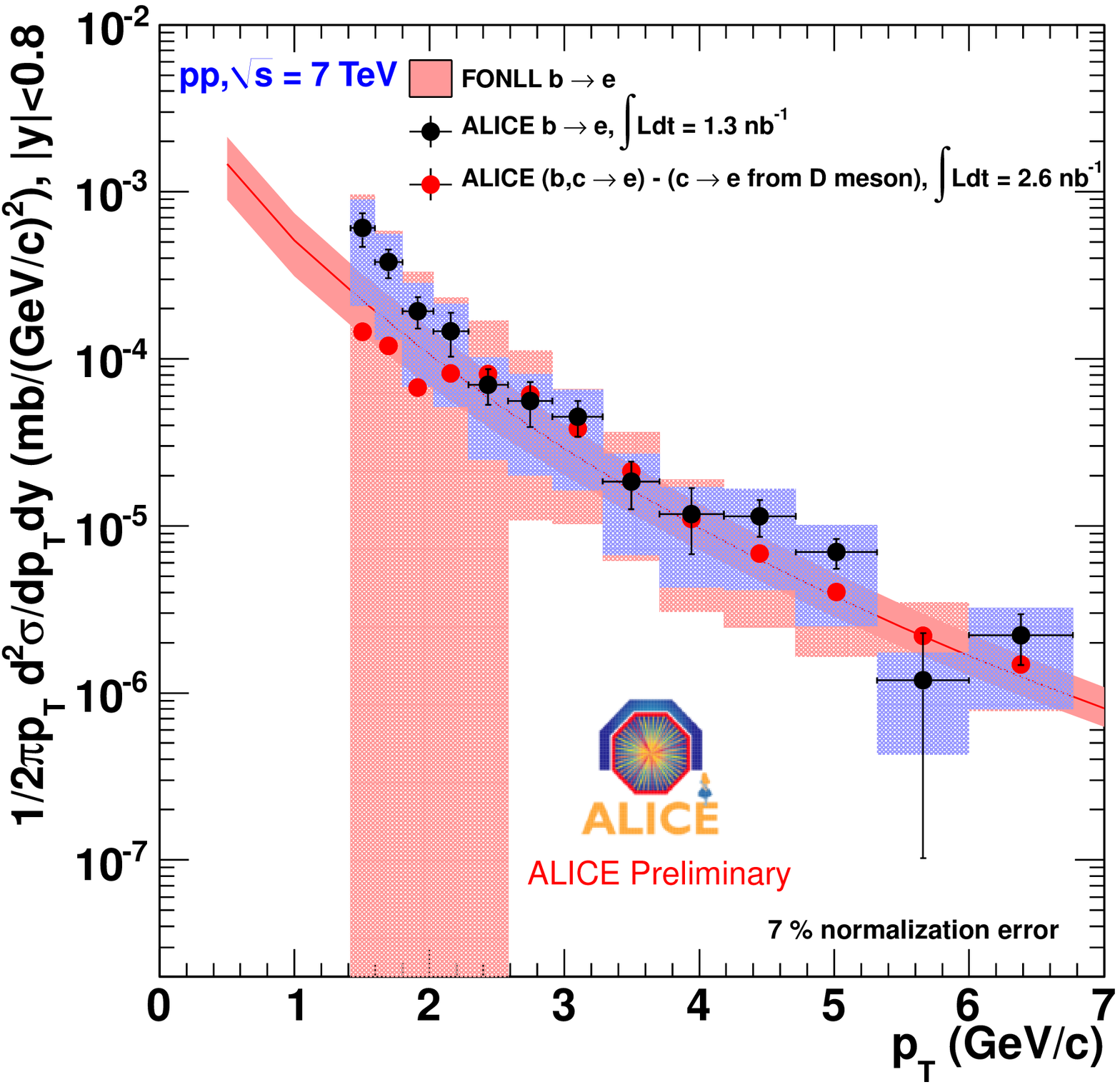}
\caption{Differential production cross sections of electrons from heavy-flavour
and D-meson decays~\cite{silvia_qm} (left panel) and of electrons from beauty 
decays (right panel) in pp collisions at $\sqrt{s} = 7$~TeV in comparison 
with FONLL pQCD calculations~\cite{fonll}.}
\label{fig:e_in_pp}
\end{figure}

The \ensuremath{p_{\rm t}}-differential production cross section of electrons 
from heavy-flavour decays is shown in the left panel of Fig.~\ref{fig:e_in_pp},
which also depicts the contribution from charm decays. 
The latter was obtained from the measured
D-meson spectra shown in Fig.~\ref{fig:d_in_pp} applying PYTHIA decay
kinematics~\cite{pythia} to electrons~\cite{silvia_qm}. FONLL calculations are 
compared to both measured electron cross sections. The agreement between the 
calculation and the data is reasonable within systematic uncertainties. Towards
low \ensuremath{p_{\rm t}}, the electron signal from heavy-flavour decays 
relative to the background from other sources becomes comparable to the 
combined systematic uncertainties of the inclusive electron measurement and the
calculated background. After background subtraction, this gives rise to a 
$\ge 100 \%$ systematic uncertainty of the electron signal from heavy-flavour 
decays at low \ensuremath{p_{\rm t}}.

\begin{figure}
\includegraphics[width=0.9\linewidth]{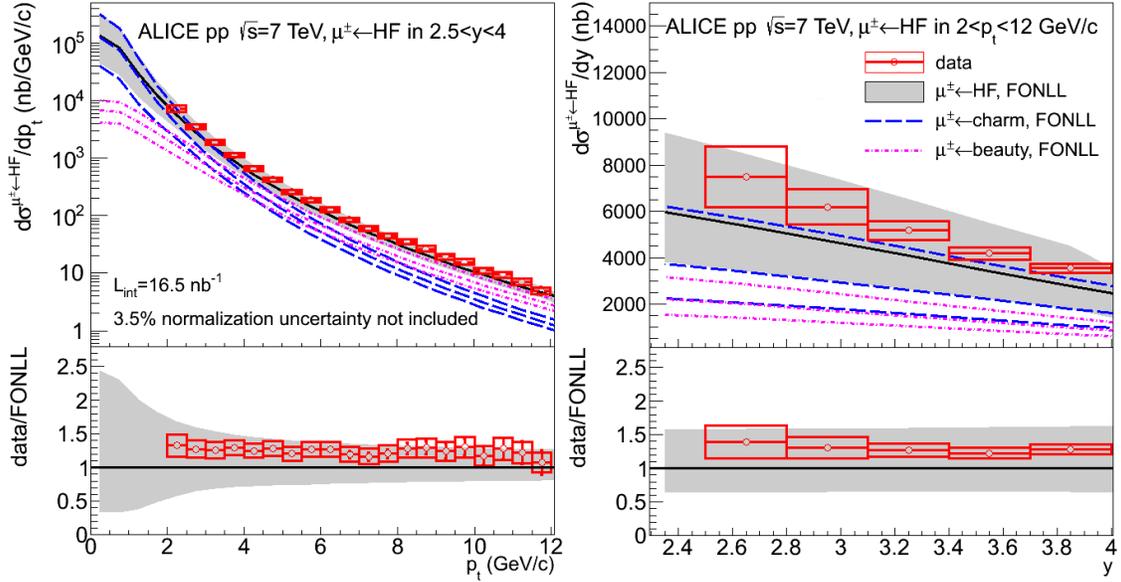}
\caption{Production cross section of muons from heavy-flavour decays in the
range $-4 < y < -2.5$ as function of \ensuremath{p_{\rm t}} and $\eta$ in pp 
collisions at $\sqrt{s} = 7$~TeV~\cite{muon_pp} in comparison with FONLL pQCD 
calculations~\cite{fonll}.}
\label{fig:mu_in_pp}
\end{figure}

The production cross section of electrons from beauty decays can be measured
using the ITS. Due to the rather long life time of beauty hadrons 
(\mbox{$c\tau($B$^{0}) = 457 \mu$m}, \mbox{$c\tau($B$^{+}) = 491 \mu$m}), 
electron tracks originating from the decay of beauty hadrons do not point back 
to the primary collision vertex. Therefore, requiring a minimum displacement of 
electron tracks from the collision vertex significantly enhances the beauty 
decay contribution to the electron sample. The remaining contribution from 
charm decays can be estimated from the D-meson cross section and it is 
subtracted~\cite{silvia_qm}. The resulting measured production cross section 
of electrons from beauty decays is shown in the right panel of 
Fig.~\ref{fig:e_in_pp}. 
A cross check is provided by the difference between the cross sections of 
electrons from heavy-flavour decays and the charm contribution (left panel of 
Fig.~\ref{fig:e_in_pp}). These two measurements of electrons from beauty
decays agree with each other within systematic uncertainties as demonstrated
in the right panel of Fig.~\ref{fig:e_in_pp}. Again a FONLL pQCD calculation
is in reasonable agreement with the measurement.

The production cross section of muons from heavy-flavour decays at forward
rapidity ($-4 < y < -2.5$) in pp collisions at $\sqrt{s} = 7$~TeV is shown 
as function of \ensuremath{p_{\rm t}} and $\eta$ in Fig.~\ref{fig:mu_in_pp} in 
comparison with results from FONLL pQCD calculations~\cite{muon_pp}. As 
observed in the other heavy-flavour channels, FONLL is in reasonable agreement 
with the data. 

In both semileptonic decay channels, i.e. electrons and muons, the data and
FONLL pQCD calculations indicate that for $p_{\rm t} > 5-6$~GeV/$c$ the dominant
contribution is from beauty decays while at lower \ensuremath{p_{\rm t}} charm 
is the relevant lepton source.

\section{Heavy-flavour pp reference at \mbox{$\sqrt{s} = 2.76$~TeV}}

The heavy-flavour production cross sections measured in pp collisions at 
$\sqrt{s} = 7$~TeV have to be scaled to $\sqrt{s} = 2.76$~TeV in order to 
provide a reference for the Pb-Pb data at the same energy per nucleon-nucleon
pair. Since FONLL pQCD calculations are in reasonable agreement with all 
heavy-flavour observables measured in pp collisions at $\sqrt{s} = 7$~TeV,
FONLL was used for the necessary $\sqrt{s}$ scaling~\cite{reference}. The 
scaling factors for D mesons, electrons, and muons were defined as the ratios 
of the corresponding cross sections from FONLL calculations at 2.76 and 7~TeV, 
where it was assumed that neither the factorisation and renormalisation scales
nor the heavy quark masses used in the FONLL calculation vary with $\sqrt{s}$. 
To evaluate the uncertainties of the scaling factors the scales and heavy quark 
masses were varied and the envelopes of the resulting scaling factors were 
determined. For all heavy-flavour observables the uncertainties of the scaling 
functions are similar. They range from $\sim 25$\% at $p_{\rm t} = 2$~GeV/$c$ to
less than 10\% for $p_{\rm t} > 10$~GeV/$c$.
${\rm D}^0$ and ${\rm D}^+$ differential production cross sections were
measured, with limited statistics, in pp collisions at 
$\sqrt{s} = 2.76$~TeV. These data are in agreement within uncertainties
with the reference obtained via $\sqrt{s}$ scaling from the 7~TeV pp 
measurements.

\begin{figure}
\includegraphics[width=0.43\textwidth]{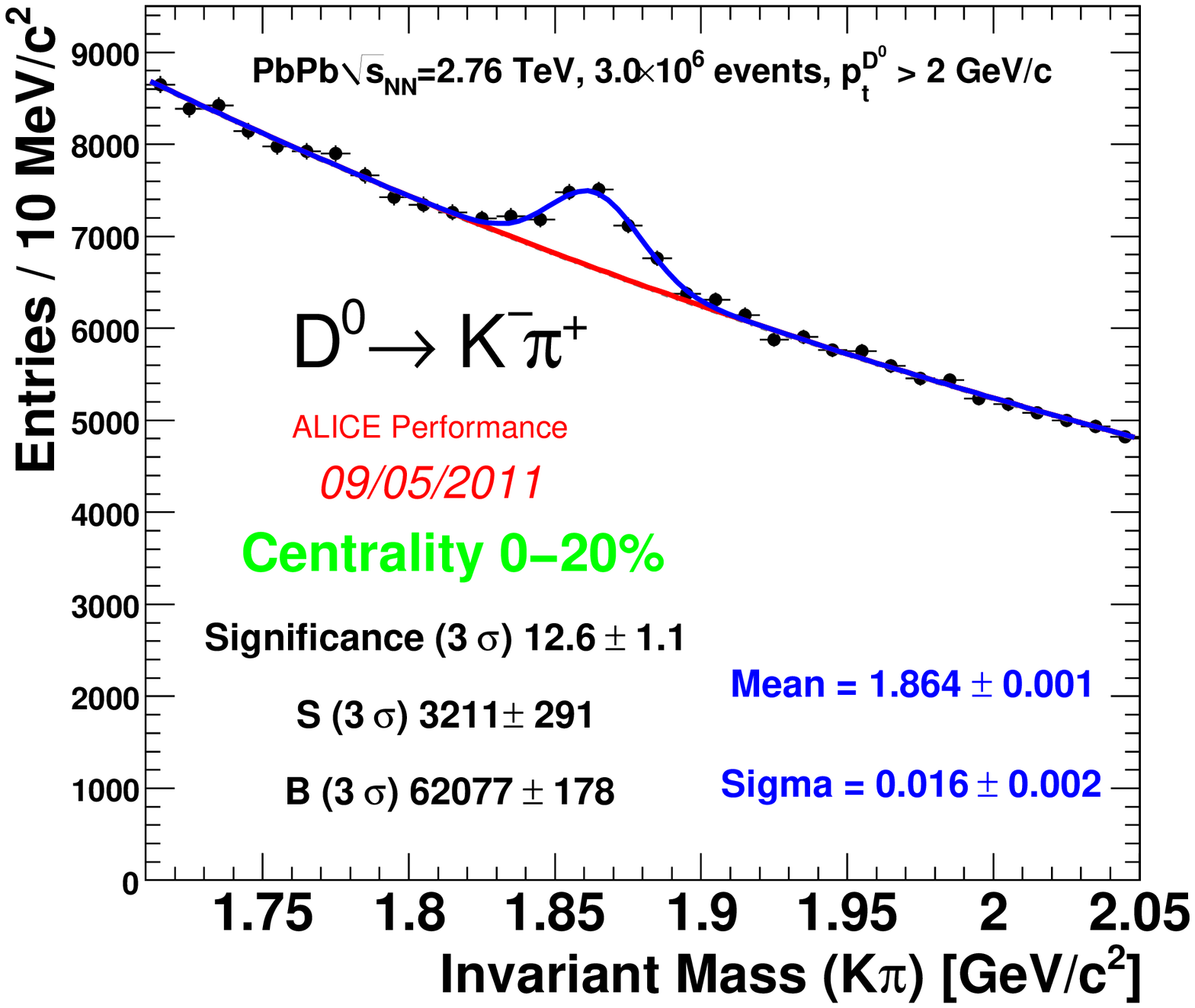}
\includegraphics[width=0.56\linewidth]{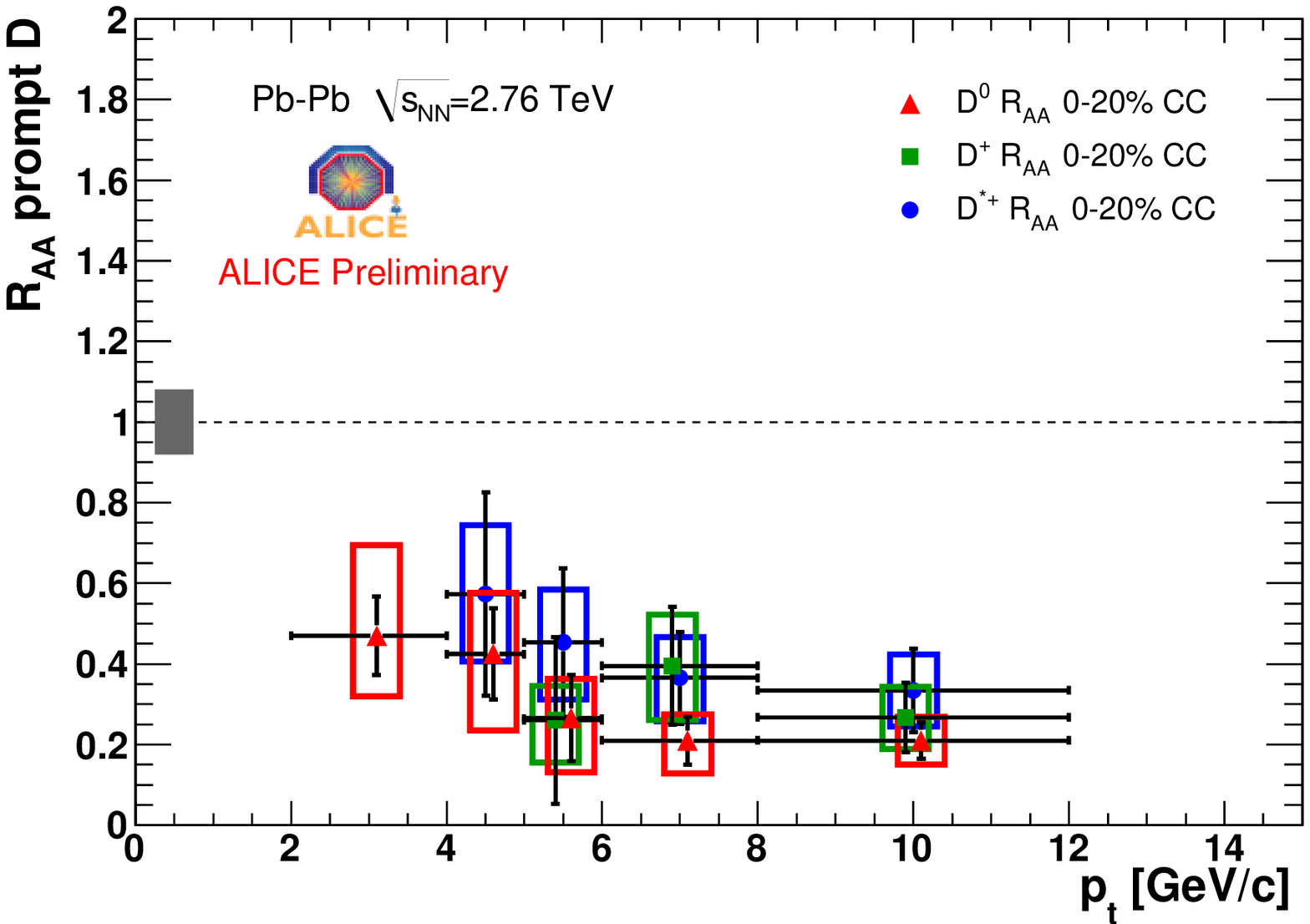}
\caption{K$\pi$ invariant mass distribution in the 20\% most central Pb-Pb 
collisions showing the ${\rm D}^0 \rightarrow {\rm K}^- \pi^+$ signal (left
panel). Nuclear modification factor \ensuremath{R_{\rm AA}} for ${\rm D}^0$, 
${\rm D}^+$, and ${\rm D}^{*+}$ mesons in central Pb-Pb collisions (right panel).
Error bars show the statistical errors. Empty and full boxes depict the 
systematic and normalisation uncertainties, respectively.}
\label{fig:d_in_pbpb}
\end{figure}

\section{Heavy flavour in Pb-Pb collisions at 
         \mbox{$\sqrt{s_{\rm NN}} = 2.76$~TeV}}

The excellent vertexing precision and particle identification capabilities
of the ALICE detector allow for D-meson reconstruction even in central
Pb-Pb collisions. As an example the reconstruction of ${\rm D}^0$ mesons
in the K$\pi$ decay channel in the 20\% most central Pb-Pb collisions is 
shown in the left panel of Fig.~\ref{fig:d_in_pbpb}. The reconstruction
efficiency is centrality independent and rises from 1 to 10\% at high 
\ensuremath{p_{\rm t}} as was determined from detailed detector simulations. 
Feed down from B decays (10-15\% after selection cuts) was subtracted and, 
since this was determined relying on FONLL calculations both for pp and Pb-Pb
collisions, the associated systematic uncertainty cancels at least partially
in the nuclear modification factor \ensuremath{R_{\rm AA}}. However, the unknown
\ensuremath{R_{\rm AA}} of B mesons introduces an additional systematic 
uncertainty on the prompt D-meson \ensuremath{R_{\rm AA}}. 
The resulting uncertainty of less than 15\% was estimated based on the 
conservative assumption that the nuclear modification factor of B mesons 
relative to the prompt D-meson \ensuremath{R_{\rm AA}} fulfils the condition 
$0.3 < R^B_{\rm AA}/R^D_{\rm AA} < 3$. The nuclear modification factors of prompt
${\rm D}^0$, ${\rm D}^+$, and ${\rm D}^{*+}$ mesons in the 20\% most central 
Pb-Pb collisions is shown in the right panel of Fig.~\ref{fig:d_in_pbpb}.
A strong suppression is observed for all species reaching a factor 3-5 for
$p_{\rm t} > 8$~GeV/$c$.

\begin{figure}
\includegraphics[width=0.49\linewidth]{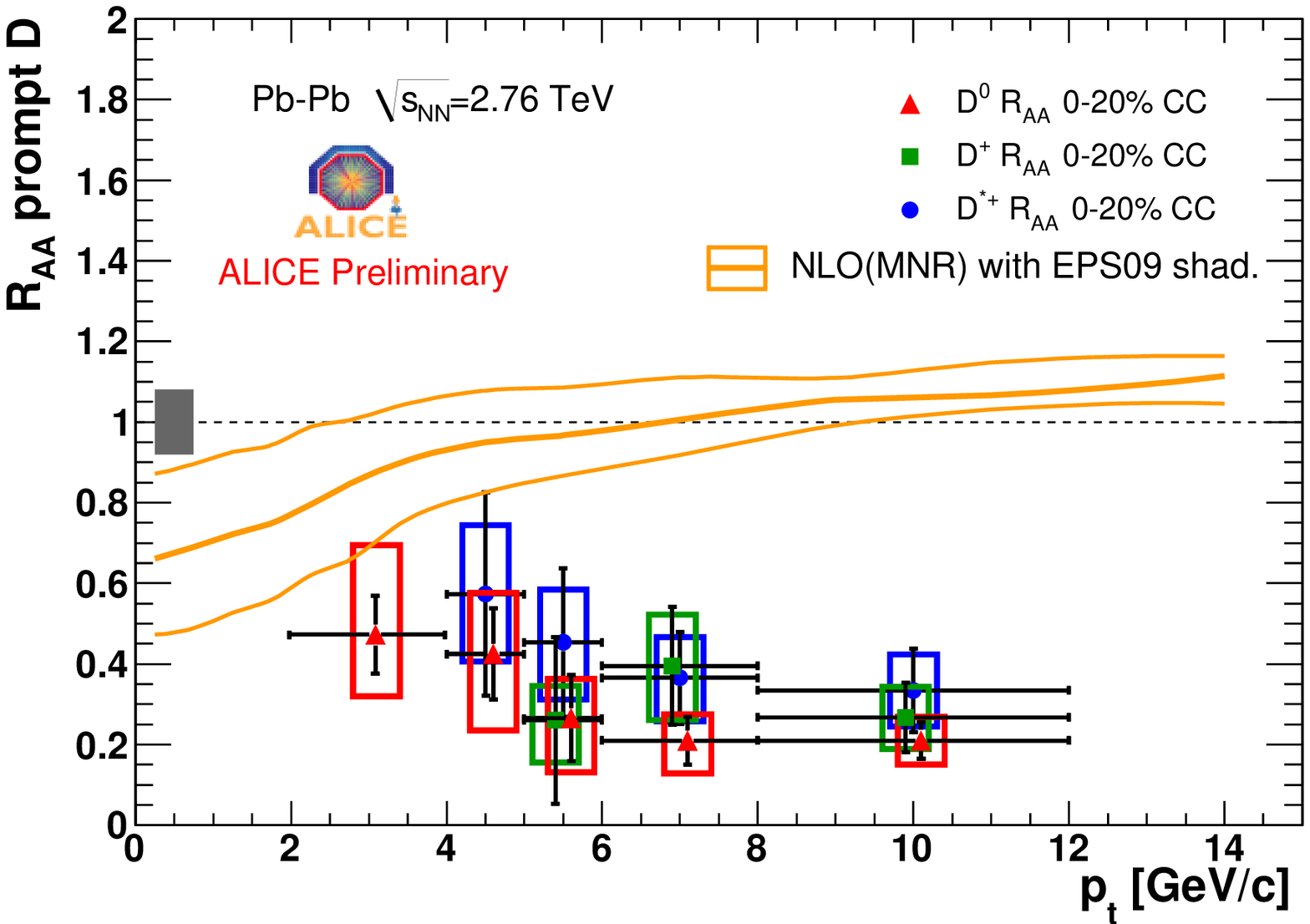}
\includegraphics[width=0.49\linewidth]{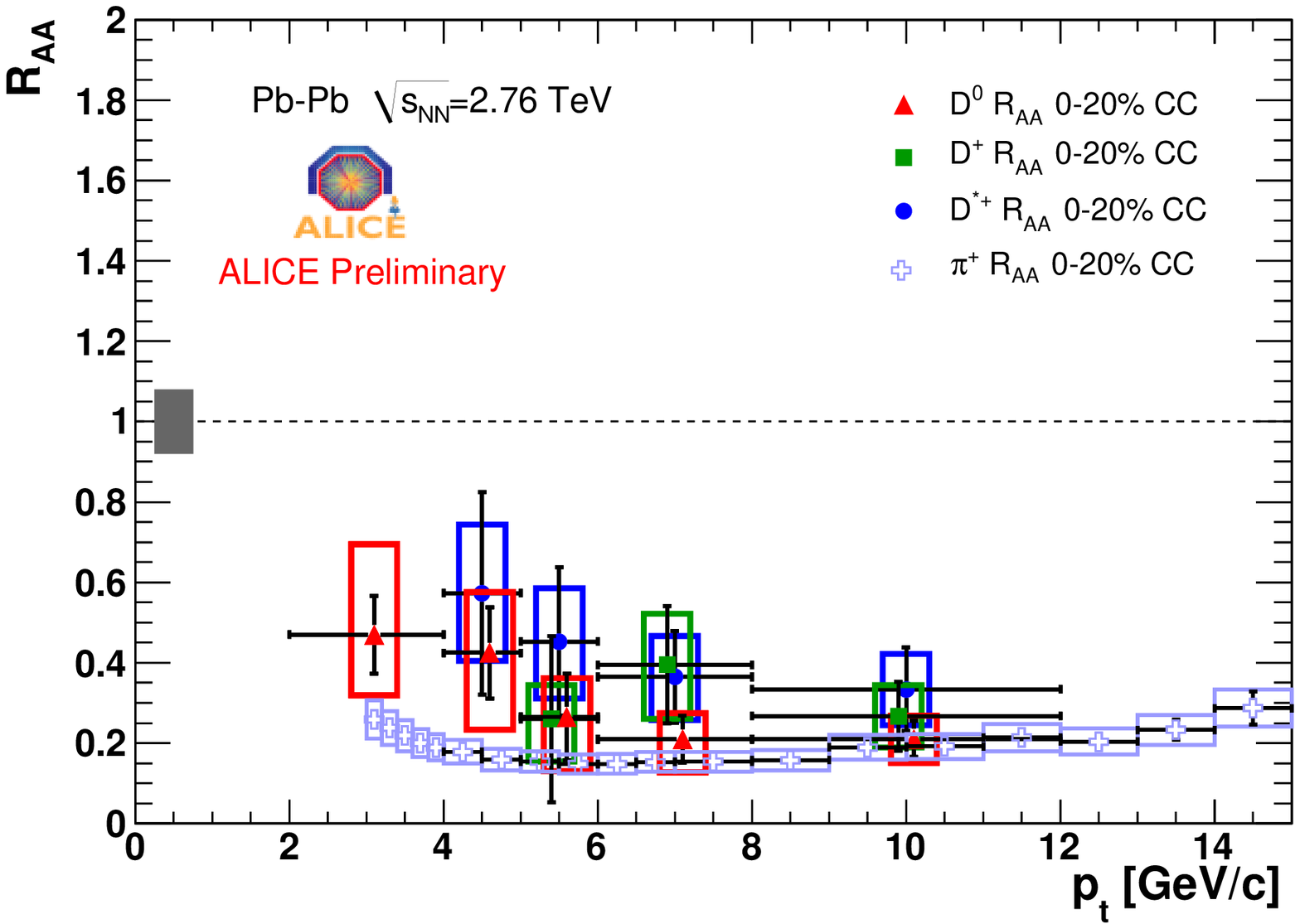}
\caption{Comparison of the measured D-meson \ensuremath{R_{\rm AA}} in central 
Pb-Pb collisions with results from an NLO pQCD calculation including EPS09 
shadowing~\cite{eps09} (left panel) and in comparison with the nuclear 
modification factor measured for charged pions (right panel).}
\label{fig:d_raa_comparisons}
\end{figure}

It is an interesting question whether the observed suppression of D mesons at
high \ensuremath{p_{\rm t}} might be related to shadowing of the parton 
distribution functions at the LHC. 
The comparison of an NLO pQCD calculation including the state of 
the art EPS09 parametrisation of parton shadowing~\cite{eps09} with the 
measured D-meson \ensuremath{R_{\rm AA}} shown in the left panel of 
Fig.~\ref{fig:d_raa_comparisons} demonstrates 
that this is not the case. Parton shadowing is not relevant for the 
interpretation of the D-meson \ensuremath{R_{\rm AA}}, i.e. the observed 
suppression clearly is a final state effect related to the hot and dense 
medium produced in Pb-Pb collisions.

Furthermore, it is important to compare the measured D-meson 
\ensuremath{R_{\rm AA}} to the nuclear modification factor of charged 
pions~\cite{raa_picharge}. This comparison is depicted in the right panel of 
Fig.~\ref{fig:d_raa_comparisons}. 
While with the current statistical and systematic uncertainties a definite 
conclusion can not be drawn the data hint towards a stronger suppression of 
pions relative to D mesons, as one would expect from a scenario in which 
induced gluon radiation is the dominant energy loss mechanism for partons 
propagating through a dense and colour-charged medium.

\begin{figure}
\includegraphics[width=0.49\textwidth]{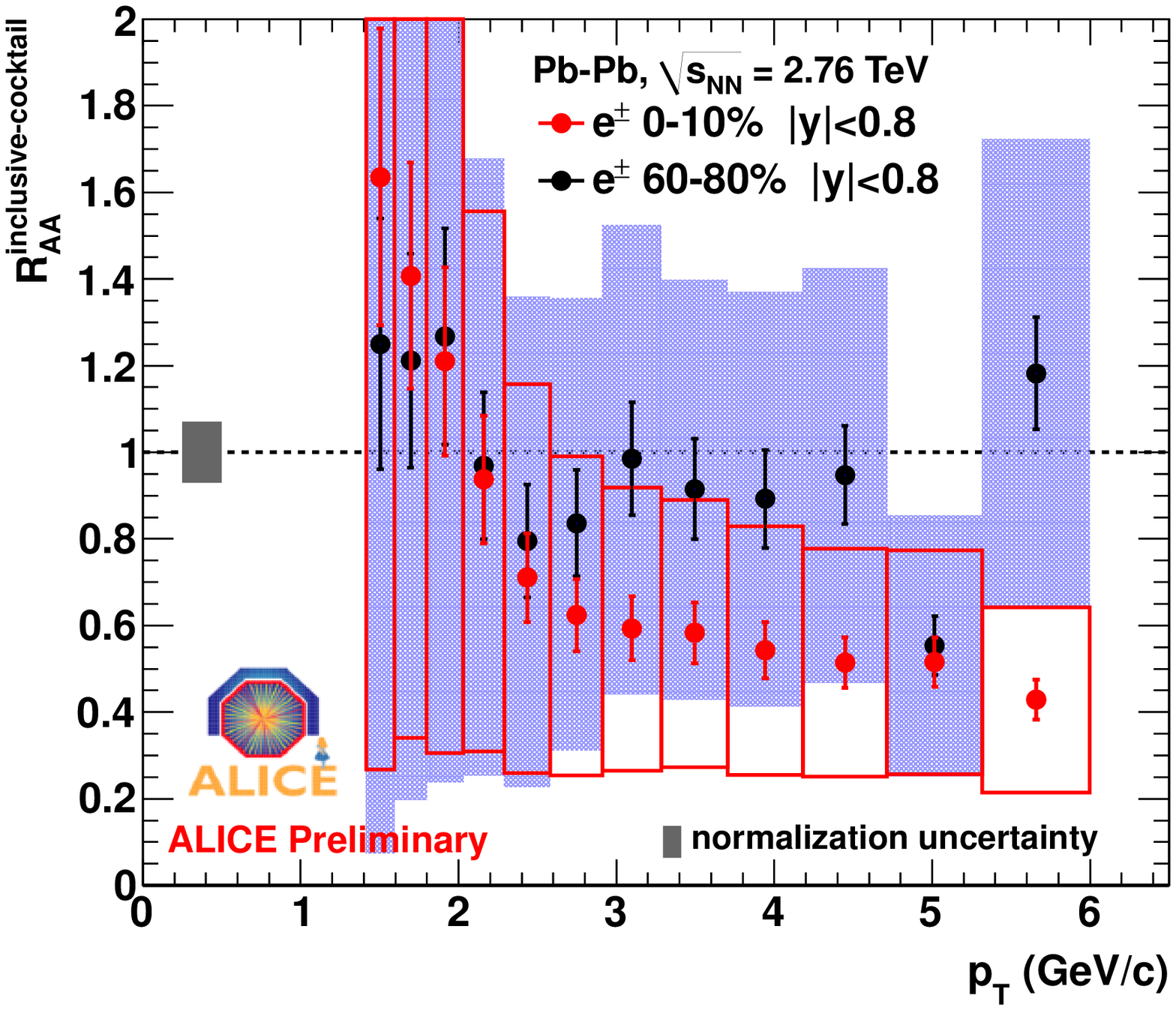}
\includegraphics[width=0.49\linewidth]{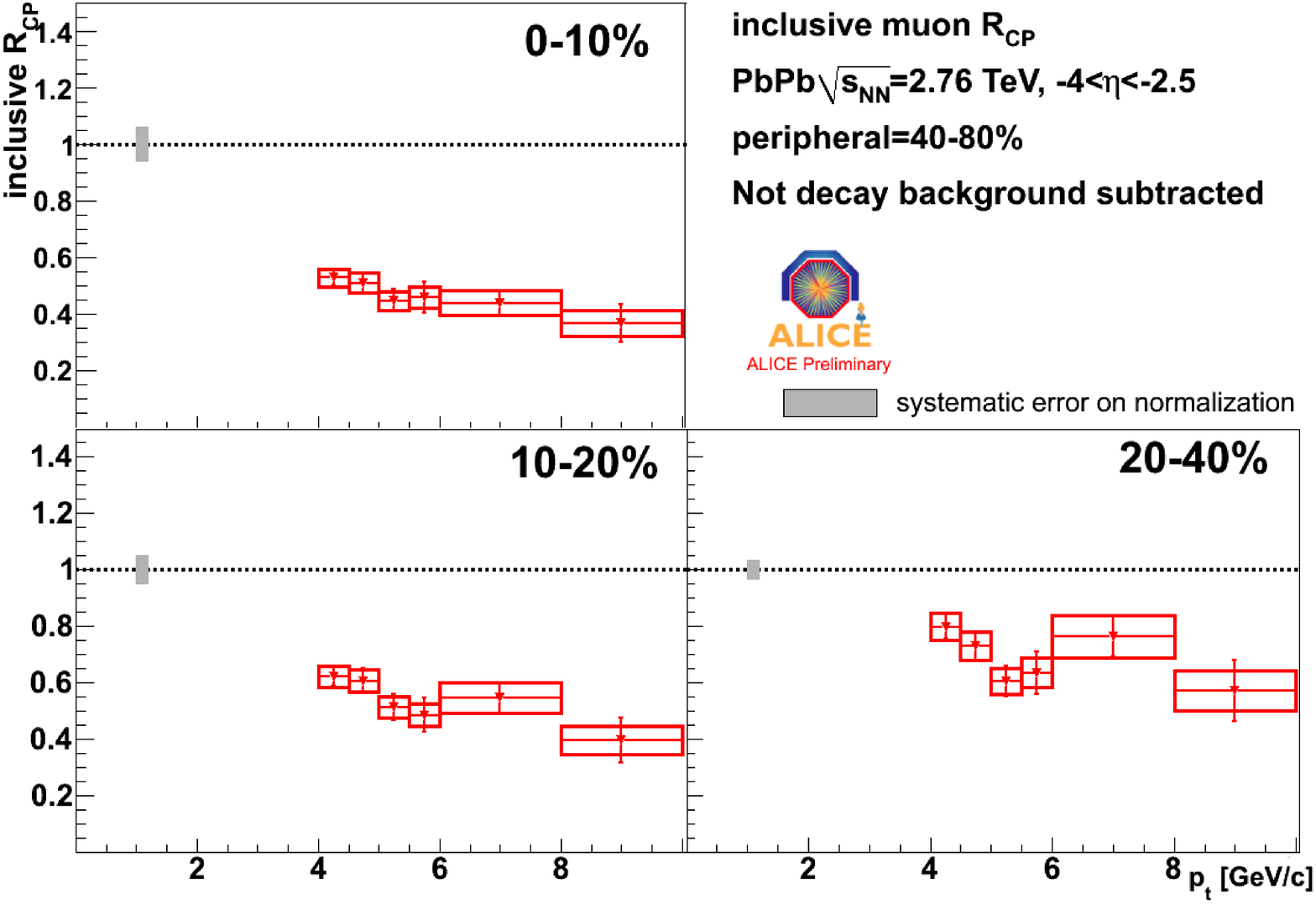}
\caption{\ensuremath{R_{\rm AA}} of cocktail subtracted electrons in central and
peripheral Pb-Pb collisions as function of \ensuremath{p_{\rm t}} (left panel). 
\ensuremath{R_{\rm CP}} of inclusive muons as function of \ensuremath{p_{\rm t}} 
in various Pb-Pb centrality classes (right panel).}
\label{fig:leptons_in_pbpb}
\end{figure}

The electron identification in Pb-Pb collisions does not yet include information
from the TRD. Therefore, the measured inclusive electron spectra are limited
to the \ensuremath{p_{\rm t}} range below 6~GeV/$c$. In this range the hadron 
contamination is less than 15\% and it is subtracted. The comparison of the 
inclusive electron spectra measured in six centrality classes with 
corresponding electron background cocktails shows a hint for an excess at low 
\ensuremath{p_{\rm t}} (up to $\sim 3$~GeV/$c$) which increases towards central 
collisions~\cite{hint_excess} and might be related to thermal photon emission 
from the hot medium as was observed in central \mbox{Au-Au} collisions at 
RHIC~\cite{thermal_rhic}. No indication for such an excess is observed in pp 
and peripheral Pb-Pb collisions. For \ensuremath{p_{\rm t}} above 
3-4~GeV/$c$ the background-subtracted inclusive electron spectra should be 
dominated by heavy-flavour decays. The corresponding \ensuremath{R_{\rm AA}} as 
function of \ensuremath{p_{\rm t}} is shown in the left panel of 
Fig.~\ref{fig:leptons_in_pbpb} for central and peripheral Pb-Pb collisions. 
Above 4~GeV/$c$ a strong suppression is observed in central collisions although
the systematic uncertainties are large. 

\begin{figure}
\includegraphics[width=0.32\textwidth]{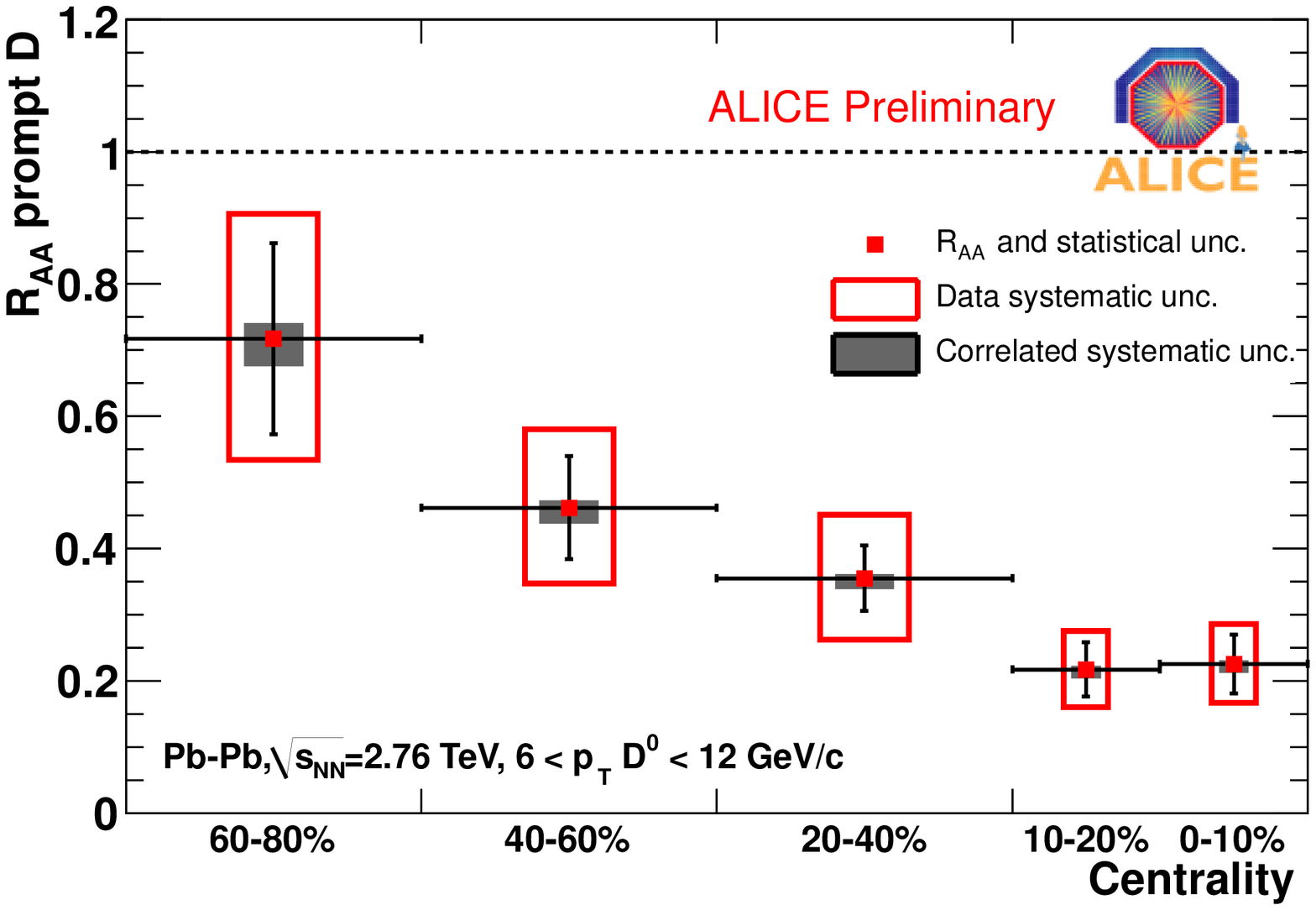}
\includegraphics[width=0.32\linewidth]{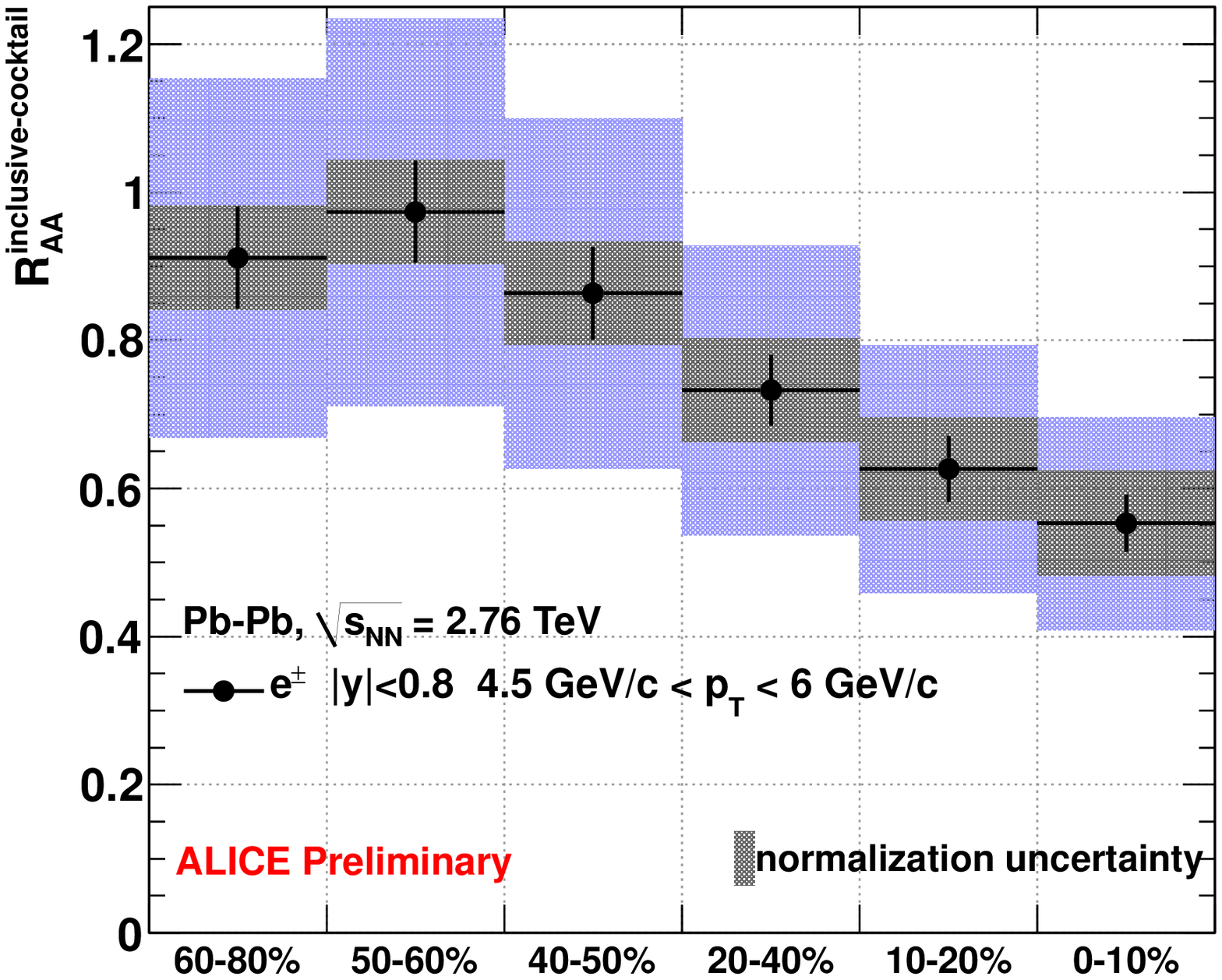}
\includegraphics[width=0.32\linewidth]{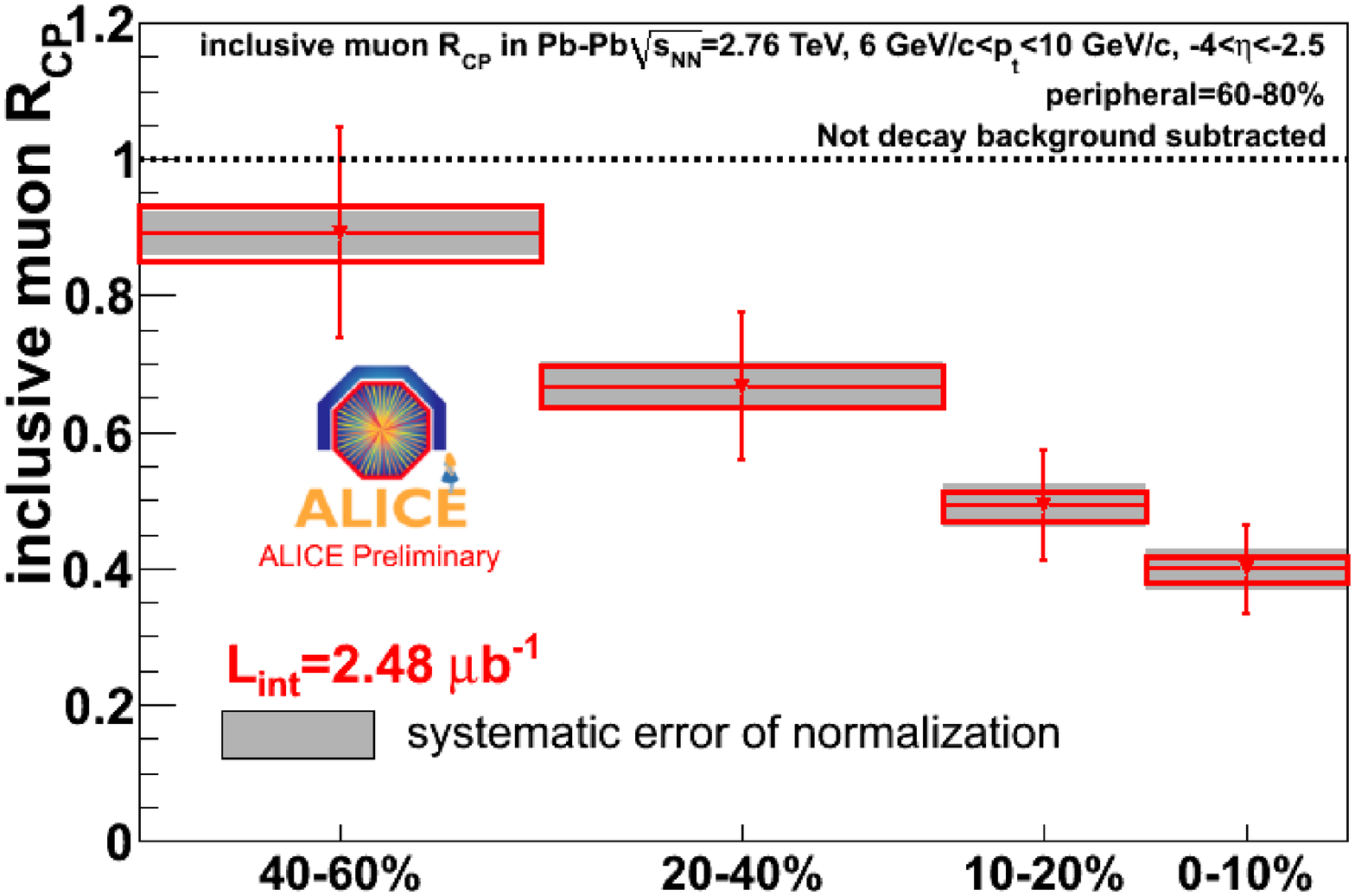}
\caption{Centrality dependence of \ensuremath{R_{\rm AA}} for prompt D mesons 
(left panel) and cocktail-subtracted inclusive electrons (middle panel) as well
as of \ensuremath{R_{\rm CP}} for inclusive muons in Pb-Pb collisions in 
different high \ensuremath{p_{\rm t}} intervals, which are indicated in the 
figures.}
\label{fig:raa_vs_centrality}
\end{figure}

The inclusive muon \ensuremath{R_{\rm CP}} is shown for various centrality bins 
as function of \ensuremath{p_{\rm t}} in the right panel of 
Fig.~\ref{fig:leptons_in_pbpb}. 
\ensuremath{R_{\rm CP}} is the ratio of yield divided by the average number of 
binary collisions in central and peripheral Pb-Pb collisions. In the inclusive 
muon \ensuremath{R_{\rm CP}} the background from light flavour decay muons was 
not subtracted. However, simulations indicate that this background is less than
10-15\% for $p_{\rm t} > 6$~GeV/$c$. A strong suppression of inclusive muons is 
observed in central relative to peripheral Pb-Pb collisions at high 
\ensuremath{p_{\rm t}}. 

As demonstrated in Fig.~\ref{fig:raa_vs_centrality} the nuclear modification
factors of prompt D mesons and cocktail-subtracted inclusive electrons as well
as \ensuremath{R_{\rm CP}} of inclusive muons at high \ensuremath{p_{\rm t}} show 
a pronounced centrality dependence.
While in peripheral Pb-Pb collisions the observed suppression is small, it
increases strongly towards central collisions.

\begin{figure}
\includegraphics[width=0.6\linewidth]{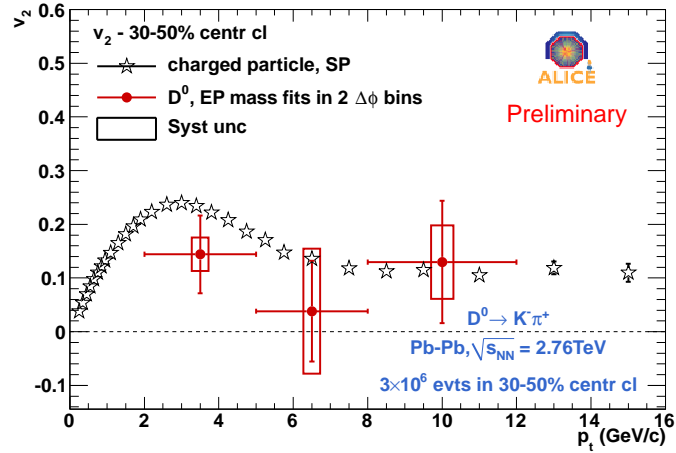}
\caption{Elliptic-flow strength $v_2$ of ${\rm D}^0$ mesons as function of 
\ensuremath{p_{\rm t}} in mid central Pb-Pb collisions in comparison with the 
corresponding unidentified charged particle $v_2$.}
\label{fig:d_flow}
\end{figure}

The measurement of nuclear modification factors of heavy-flavour probes
indicate a strong interaction of heavy quarks produced in the earliest
phase of Pb-Pb collisions at the LHC with the hot and dense partonic
medium that is formed afterwards. It is an important question whether
heavy quarks thermalise in this medium. This question can be addressed 
via the measurement of the elliptic flow strength $v_2$ of particles carrying
heavy quarks. $v_2$ is the second coefficient of the Fourier expansion of
the azimuthal distribution of particles relative to the reaction plane of
the collision, which is defined by the azimuth of the impact parameter and 
the beam direction. The fireball produced in the initial state of a 
nucleus-nucleus collision exhibits a spatial anisotropy with respect to
this reaction plane. The resulting asymmetric pressure gradients lead to
an asymmetric momentum distribution of particles in the final state, quantified
by a non-zero value of $v_2$. The elliptic flow strength $v_2$ was measured
for neutral D mesons~\cite{d0_flow} as shown in Fig.~\ref{fig:d_flow}. A 
correction for feed down from B mesons was not applied yet. While statistical 
and systematic uncertainties are still large, an indication for a non-zero 
$v_2$ of D mesons is observed.

\section{Summary}
The production of heavy flavour has been investigated with the ALICE experiment
at the CERN LHC in pp collisions at $\sqrt{s} = 7$~TeV and in Pb-Pb 
collisions at $\sqrt{s_{\rm NN}} = 2.76$~TeV. In pp collisions the differential 
production cross sections of D mesons at mid-rapidity, as well as the cross
sections for electron and muon production from semileptonic heavy-flavour 
hadron decays at mid- and forward-rapidity, respectively, have been measured. 
State of the art FONLL pQCD calculation are in agreement with all of these 
measurements within statistical and systematic uncertainties. In Pb-Pb 
collisions nuclear modification factors have been measured for D mesons 
as well as electrons and muons from heavy-flavour decays. A suppression 
indicating the energy loss suffered by heavy quarks while they propagate 
through the medium produced in central Pb-Pb collisions is observed. 
In addition, a hint for non-zero elliptic flow of neutral D mesons might 
indicate that charm quarks participate in the collective dynamics of this
strongly coupled partonic medium. 

Additional data from both the pp and, in particular, the Pb-Pb running 
periods with ALICE in the year 2011 at the LHC, going hand in hand with
an even improved understanding of the detector response and better calibration
and reconstruction procedures, will significantly enhance the heavy-flavour
physics reach of ALICE. Not only will the \ensuremath{p_{\rm t}} reach of the 
observables presented in this writeup be increased considerably, both to lower 
and to higher values of \ensuremath{p_{\rm t}}, but also new observables will 
become accessible, e.g. baryons carrying charm \mbox{$(\Lambda_c)$}, 
heavy-flavour correlations, and beauty jets.

%
% Bibliography
%

\end{document}